\documentclass[aps,prl,twocolumn,tightenlines]{revtex4-1}
\usepackage{graphicx}
\usepackage{amssymb,amsmath}
\usepackage{color}
\usepackage{bm}

\newcommand{\ignore}[1]{}

\begin{document}

\title{Exact statistical mechanics of the Ising model on networks}
% Exact Ising partition function computed on {sparse, tree-like} network {with small tree-width} 
% augmented, complex, enhanced, generalized
\author{Konstantin Klemm}
\affiliation{IFISC (CSIC-UIB), Palma de Mallorca, Spain}

\date{\today}
\begin{abstract}
The Ising model is an equilibrium stochastic process used as a model in several branches of science including magnetic materials \cite{Ising:1925}, geophysics \cite{MaNJP:2019}, neuroscience \cite{SchneidmanNature:2006}, sociology \cite{StaufferAJP:2008} and finance \cite{Bornholdt:2001}.
Real systems of interest have finite size and a fixed coupling matrix exhibiting quenched disorder. Exact methods for the Ising model, however, employ infinite size limits, translational symmetries of lattices and the Cayley tree \cite{BaxterBook:1989}, or annealed structures as ensembles of networks \cite{DorogovtsevPRE:2002}. Here we show how the Ising partition function can be evaluated exactly by exploiting small tree-width. This structural property is exhibited by a large set of networks \cite{KlemmJPC:2020}, both empirical and model generated.
\end{abstract}

\maketitle

%%% FIGURE 1, ILLUSTRATIONS OF 2-TREES AND KARATE CLUB COMBINED

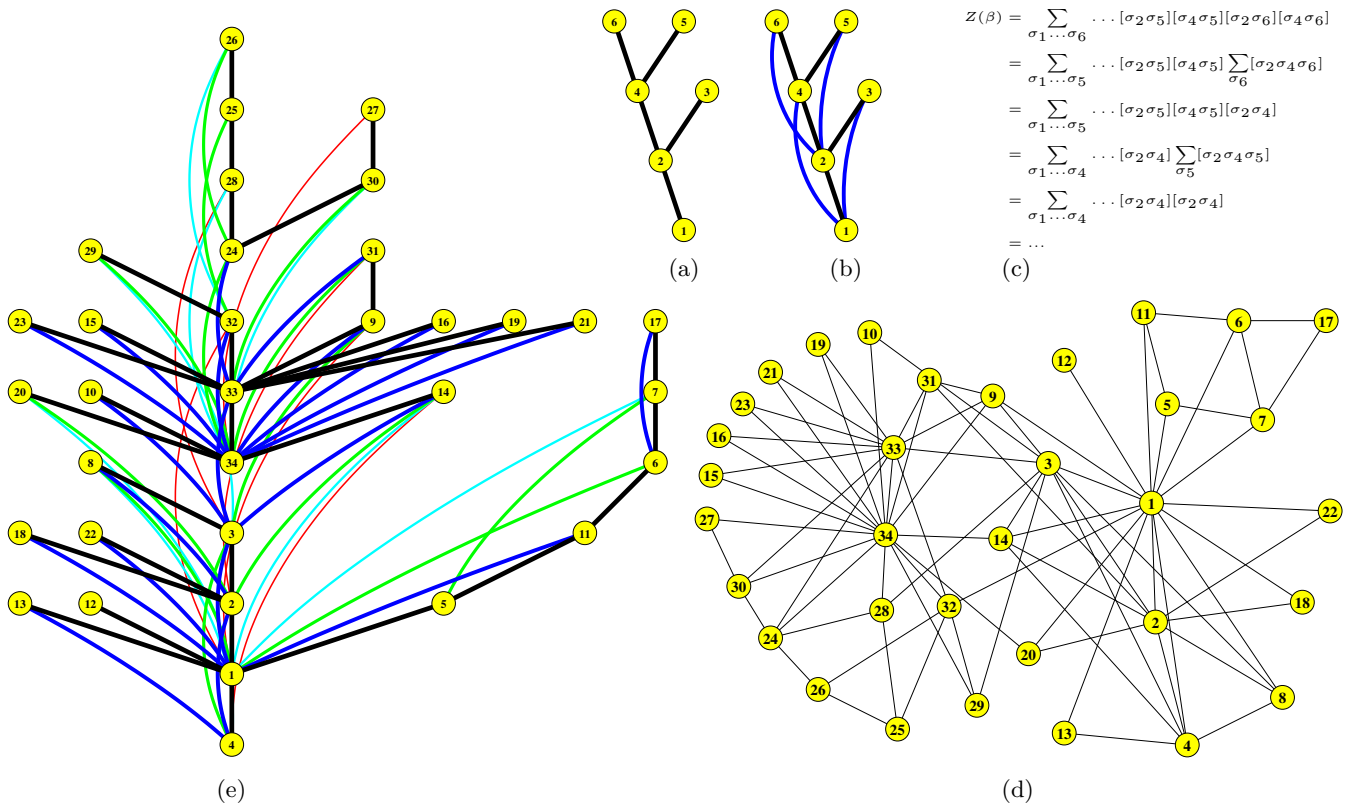
\begin{figure*}
\begin{picture}(500,310)
\put(0,20){\includegraphics[width=250pt]{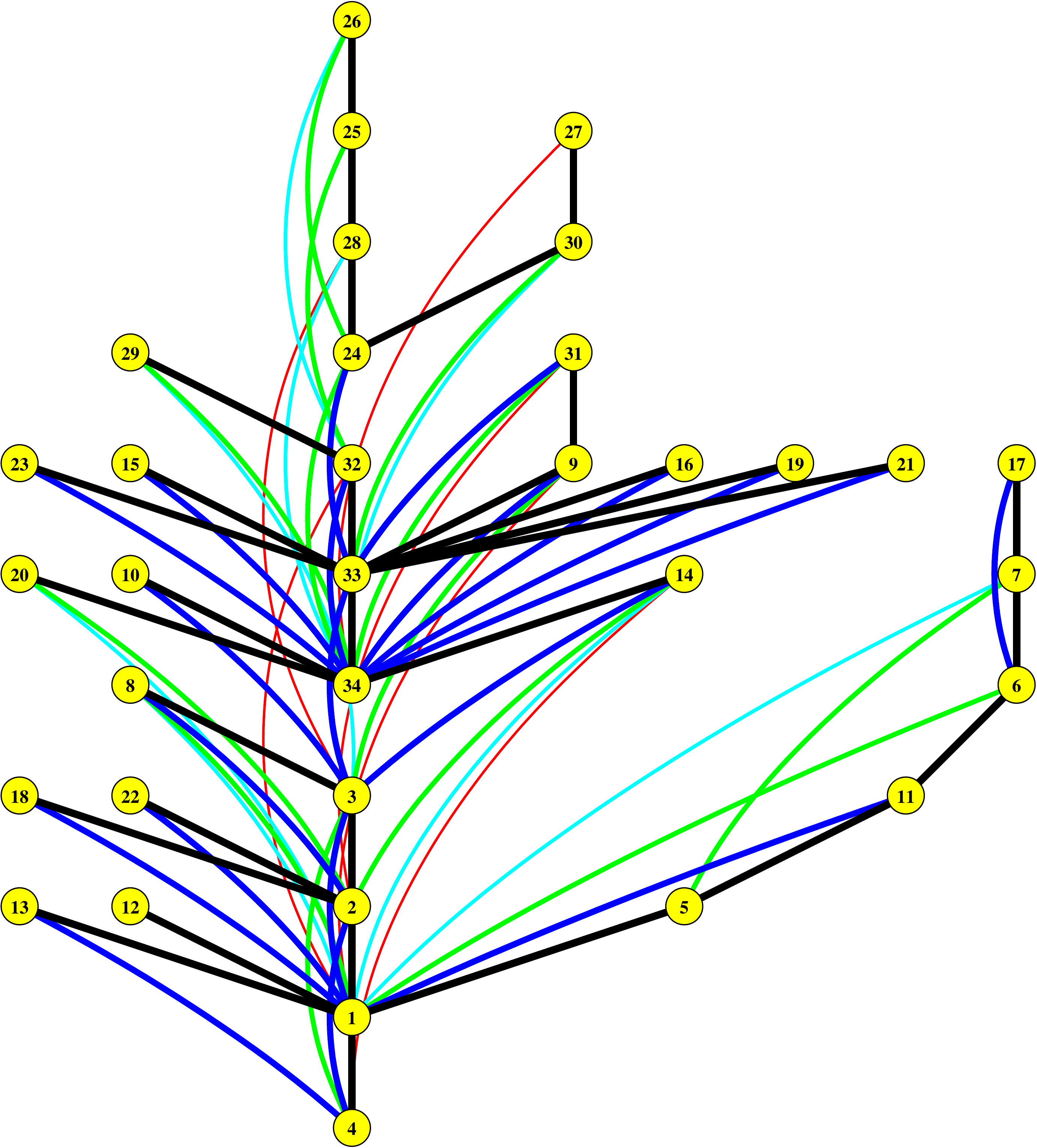}}
\put(79,5){(e)}
\put(260,20){\includegraphics[width=245pt]{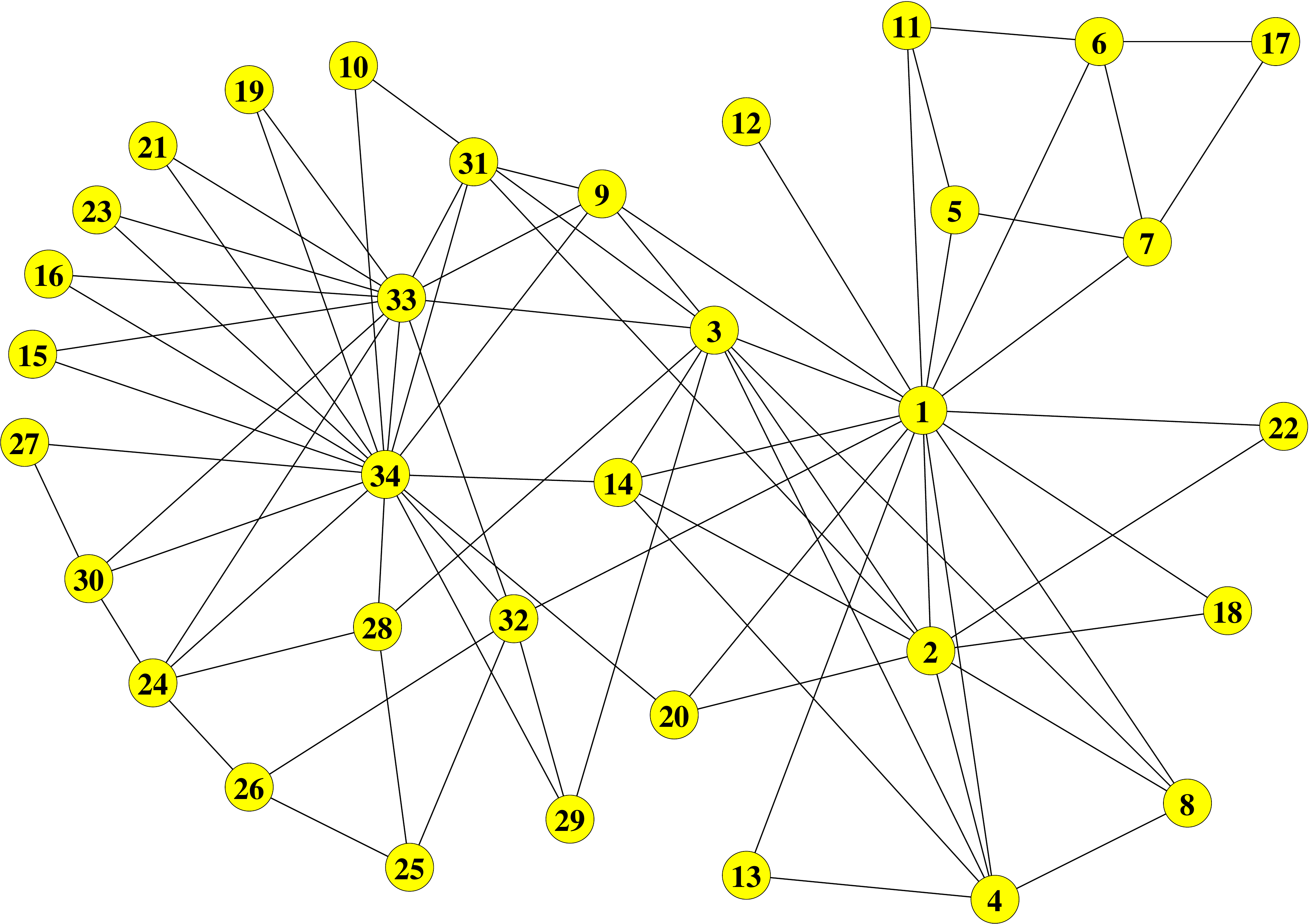}}
\put(376,5){(d)} 
\put(225,215){\includegraphics[height=31mm]{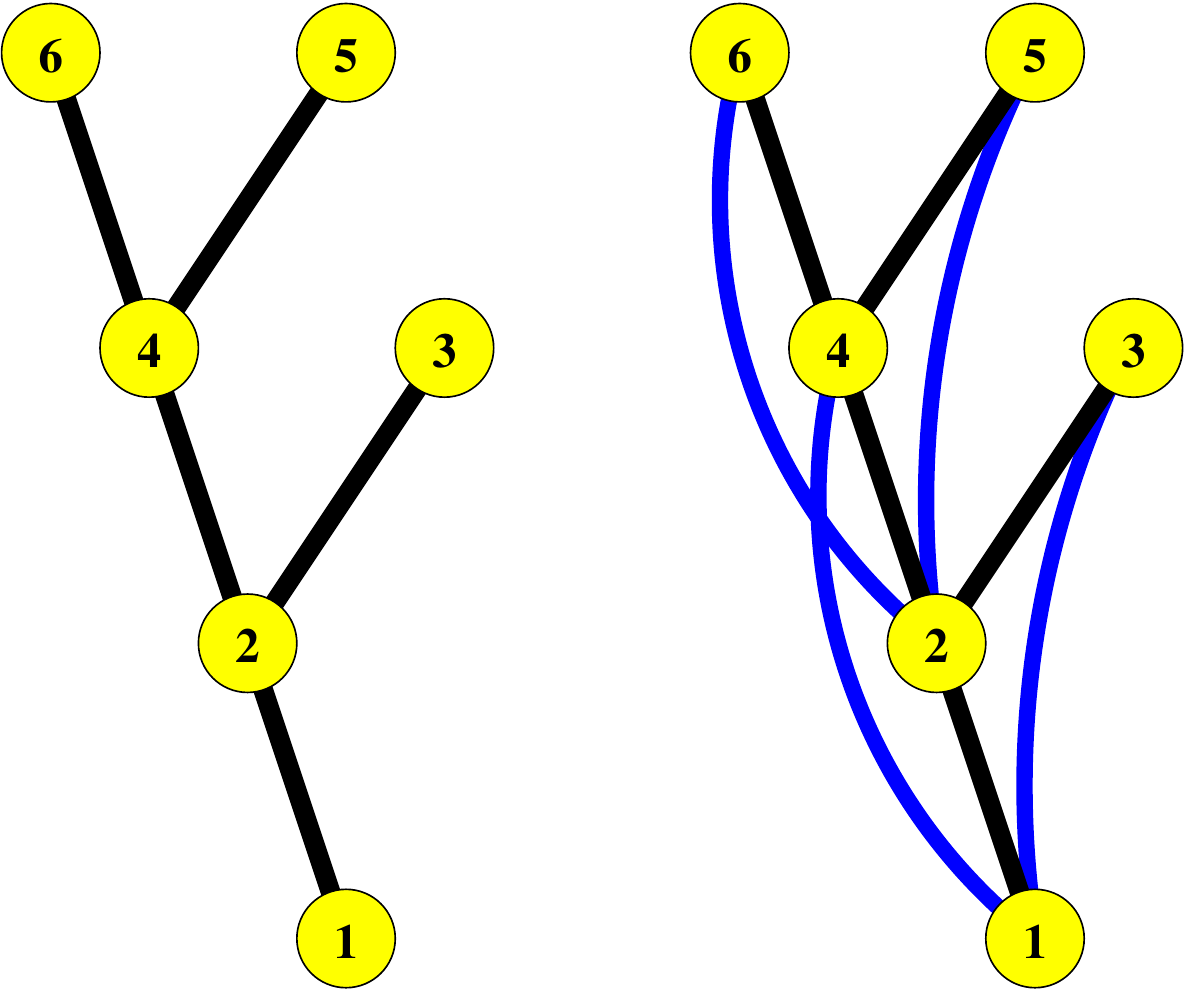}}
\put(250,202){(a)}
\put(311,202){(b)}
\put(360,260){\begin{minipage}{5cm}
\tiny 
\begin{align*} 
Z(\beta) & = \sum_{\sigma_1 \dots \sigma_6} \dots [\sigma_2 \sigma_5] [\sigma_4 \sigma_5] [\sigma_2 \sigma_6] [\sigma_4 \sigma_6]\\
         & = \sum_{\sigma_1 \dots \sigma_5} \dots [\sigma_2 \sigma_5] [\sigma_4 \sigma_5] \sum_{\sigma_6} [\sigma_2 \sigma_4 \sigma_6] \\
         & = \sum_{\sigma_1 \dots \sigma_5} \dots [\sigma_2 \sigma_5] [\sigma_4 \sigma_5] [\sigma_2 \sigma_4] \\
         & = \sum_{\sigma_1 \dots \sigma_4} \dots [\sigma_2 \sigma_4] \sum_{\sigma_5} [\sigma_2 \sigma_4 \sigma_5]\\
         & = \sum_{\sigma_1 \dots \sigma_4} \dots [\sigma_2 \sigma_4] [\sigma_2 \sigma_4]\\
         & = ...
\end{align*}
\end{minipage}
}
\put(376,202){(c)}

\end{picture}
\caption{\label{fig:illustrations}
Complete and partial $k$-trees and the Ising model. (a) A tree rooted at the bottom (node 1). (b) From the tree of panel (a), a 2-tree is obtained by adding a bond (shown in blue) from each node to its parent's parent node. (c) Computing the Ising partition function on the 2-tree shown in (d): a term $[\dots]$ stands for an arbitrary function of the spin variables within brackets. In the initial expression, there is one such function of the form $\exp(\beta \sigma_v \sigma_w)$ for each bond $\{v,w\}$. (d) Karate club network, see Methods for details. (e) The Karate club network is a partial $5$-tree. The 34 nodes of the network are arranged in a tree rooted at node 4. Each bond $\{v,w\}$ of the network is parallel to this tree, so $v$, $w$ and the root all lie on one tree path; and the distance between $v$ and $w$ is indicated by the colour of the bond, being red for the maximum value $5$. Based on this representation, the computation of the partition function looks at most 5 spin variables ahead.
}
\end{figure*}

%%% TASK EXPOSITION AND GENERAL SOLUTION

A network $(V,B)$ is given by a finite node set $V=\{1,2,\dots,n\}$ and bond set 
$B \subseteq \{ b \subset V : |b|=2\}$. Figure \ref{fig:illustrations}(d) is an example of a small social network. The Hamiltonion $H$ assigns each spin configuration $\sigma = (\sigma_1, \sigma_2, \dots \sigma_n) \in \{-1,+1\}^V$ an energy
\begin{equation}
H(\sigma) = -\sum_{\{v,w\} \in B} \sigma_v \sigma_w~.
\end{equation}
For the macrocanonical ensemble at temperature $T = \beta^{-1}$, our goal is to numerically evaluate the partition function
\begin{align}
Z(\beta) & = \sum_{\sigma \in \{-1,+1\}^n}\exp[-\beta H(\sigma)] \label{eq:ZwithH} \\
         & = \sum_{\sigma \in \{-1,+1\}^n} \prod_{\{v,w\} \in B}\exp[\beta \sigma_v \sigma_w] \label{eq:ZwithProd}
\end{align}
The most naive approach directly performs the sum over all $2^n$ spin configurations. Efficiency is gained by summing over spin variables in an order chosen to suit the given network. Starting from a set of factors, each dependent on exactly two spin variables as in expression (\ref{eq:ZwithProd}), the step-wise summation proceeds as follows. Let $x$ be the next node in the given order; multiply (expand) all factors dependent on $\sigma_x$, thus obtaining a single $\sigma_x$-dependent factor $F_x$; perform the summation over $\sigma_x$; repeat until summation over all spin variables is done.

%%% TREE EXAMPLE

How do we choose an order of nodes that reduces computational cost? Crucially, part (3) of the computation generates a function $F_v$ depending at least on the neighbouring spin variables of node $v$ not yet summed over. Storing $F_v$ as a table of function values, this will use $O(2^k)$ in time and memory for $F_v$ dependent on $k$ spin variables. If the network is a tree, the nodes are optimally ordered by descending distance from the root, such as node ordering $(6,5,4,\dots,1)$ in the tree shown in Figure~\ref{fig:illustrations}(a). This ensures that the multiplication before each summation generates factors dependent on at most two spin variables.

%%% TREE-LIKE APPROXIMATION, BELIEF PROPAGATION ETC

The exact and fast computation on trees forms the basis for approximative methods for general networks. In {\em belief propagation} \cite{MezardBook:2009}, bits of information (so-called belief values) are passed along the bonds of the network. At each node $x$, the incoming information is compiled assuming neighbours of $x$ being conditionally independent given spin $\sigma_x$. This approach is exact when the network is a tree. It still produces acceptable approximations when the network is locally tree-like \cite{DorogovtsevRMP:2008} so the length of cycles exceeds the expected correlation length of the process, at least in the limit of large system size \cite{DemboAAP:2010}. Most real networks, however, contain a large amount of short cycles due to their cliquishness \cite{Watts:1998} and modularity \cite{NewmanPNAS:2006}. Belief propagation has been been generalized to account for short cycles at least partially \cite{YedidiaNIPS:2000, RadicchiPRE:2016, Cantwell:2019, Kirkley:2021}.

%%% k-TREE and 2-TREES
In the present approach, we consider a different notion of tree-like structure for which exact rather than approximate computation is feasible. For integer $k\ge 1$, a $k$-tree \cite{Bodlaender:2010} is obtained by iteratively adding a node $x$ and bonds from $x$ to each node in a $k$-clique. Growth starts from a $k$-clique network as initial condition. Since a $1$-clique is a single node, a $1$-tree is just a tree. A $2$-clique is two nodes $v,w$ with a bond $\{v,w\}$. A $2$-tree is thus grown by choosing a bond $\{v,w\}$ in an existing $2$-tree and joining nodes $v,w,x$ into a triangle with a newly added node $x$. Equivalently, a $2$-tree is obtained by augmenting a tree with a bond from each child node to the parent's parent as shown in Figure \ref{fig:illustrations}(b). Using the same node ordering as for the underlying tree, the exact computation of the partition function is straight-forward, now involving factors depending on 3 spin variables at most. Figure \ref{fig:illustrations}(c) sketches an example of the computation.

%%% FIGURE 2, RESULTS FOR 2-TREES
\begin{figure}
\centerline{\includegraphics[width=0.48\textwidth]{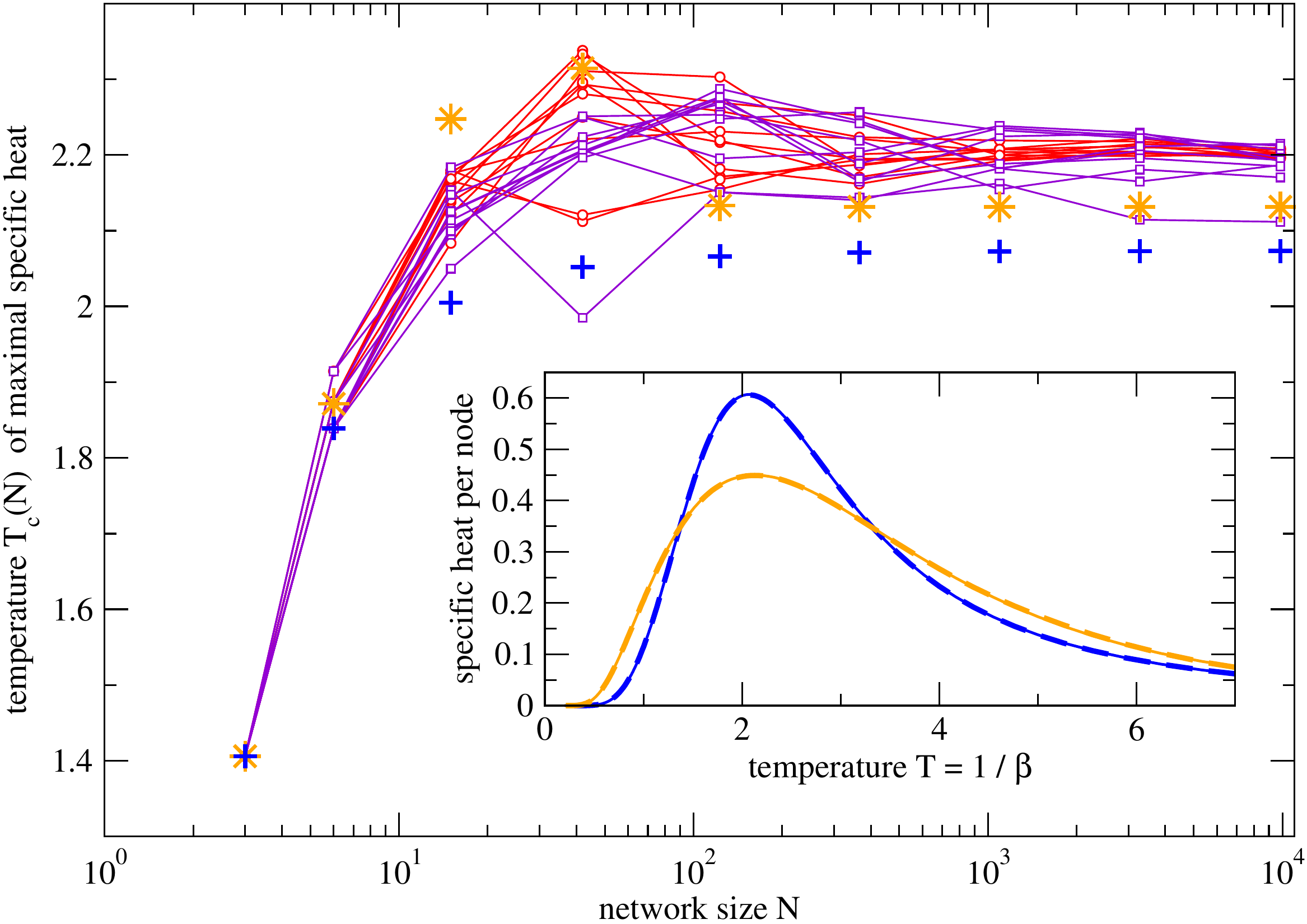}}
\caption{\label{fig:crit_twotrees}
Maxima of the specific heat on artificial 2-trees. Results for 10 realizations from the deactivation model \cite{KlemmPRE:2002a} (red lines and circles) and for 10 realizations from the stochastic triangle attachment model \cite{DorogovtsevPRE:2001} (violet lines and squares) are shown together with those for deterministic triangle attachment \cite{DorogovtsevPRE:2002b} (orange stars) and one-dimensional grids with coordination number 4 (blue crosses). For the latter two, the inset shows the temperature dependence of the specific heat per node, using the same colours as in the main panel and for sizes of $1095$ (thin dashed curves) and $9843$ nodes (thick solid  curves). The exact specific heat $C(T)$ involves the first and second derivatives of $Z(\beta)$, computed analogously to $Z(\beta)$ itself, see Methods.
}
\end{figure}

%%% RESULTS for 2-TREES
2-trees have been used as models of scale free-networks, including the sequential and parallel growth rules by Dorogovtsev et al.\ \cite{DorogovtsevPRE:2001,DorogovtsevPRE:2002b} and the model by Klemm and Egu\'{\i}luz \cite{KlemmPRE:2002a}, see Methods. Figure~\ref{fig:crit_twotrees} shows the critical Ising temperature $T_c(N)$, i.e. the value of $T$ maximizing the specific heat, for artificial 2-trees with sizes $N$ up to $10^4$ nodes. The plots suggest that $T_c(N)$ remains bounded for growing $N$, while $T_c(N) \sim \ln N$ is found for other growing and uncorrelated scale-free networks \cite{DorogovtsevRMP:2008}.

%%% FIGURE 3, HEAT CAPACITY OF EMPIRICAL NETWORKS
\begin{figure}
\centerline{\includegraphics[width=0.48\textwidth]{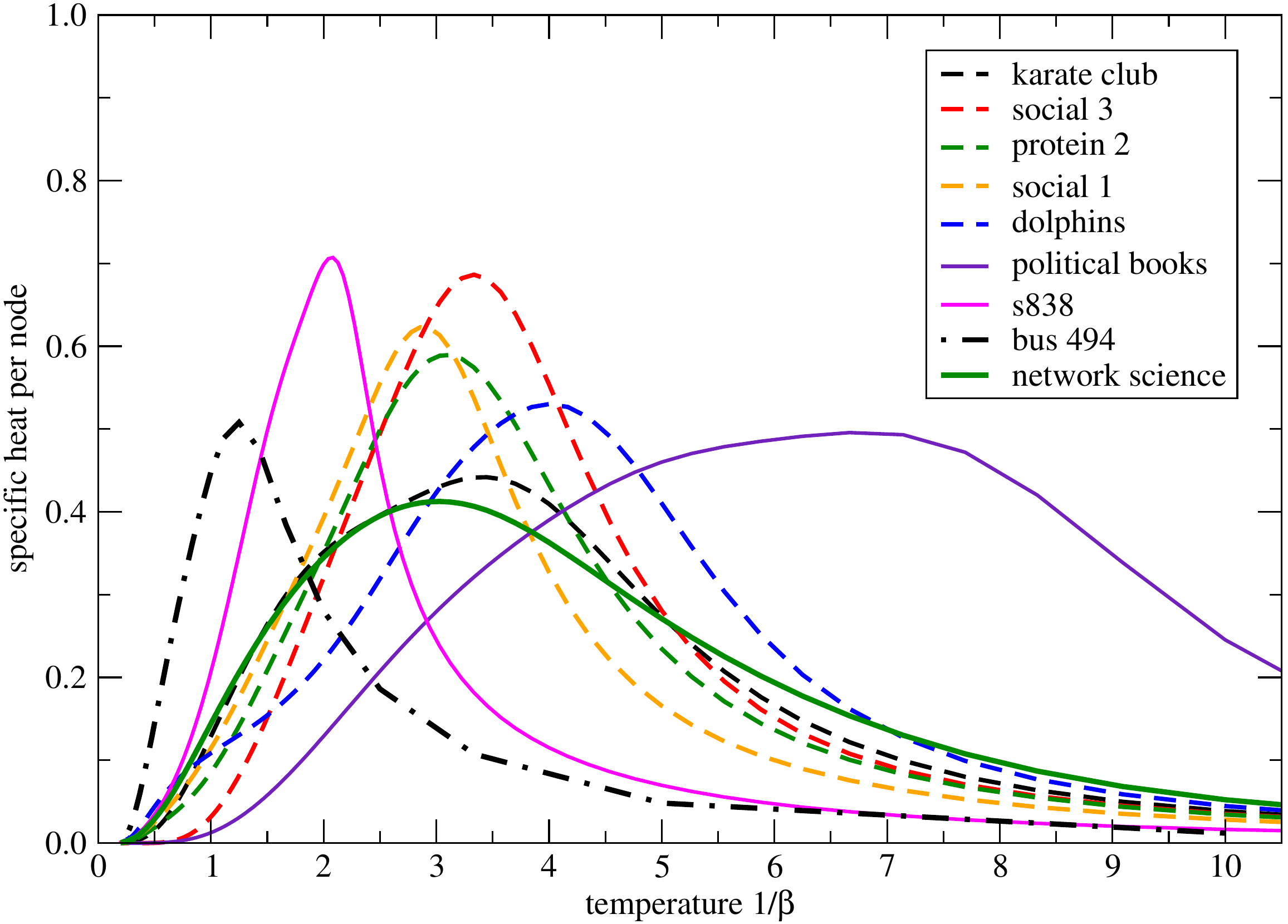}}
\caption{\label{fig:heat}
Exact specific heat of the Ising model on empirical networks. Plotted values are $C(\beta)/N$ for number of nodes $N$. Networks in the legend are ordered by number of bonds, ranging from 78 (karate club) to 914 (network science).
}
\end{figure}

%%% RESULTS FOR EMPIRICAL NETWORKS
Empirical networks, though not being $k$-trees, may be represented as {\em partial} $k$-trees, often for $k\ll n$ \cite{KlemmJPC:2020}, by keeping only a subset of bonds in a $k$-tree. Then there is an ordering of the nodes for which the calculation of the partition function involves factors depending on at most $k+1$ spin variables. The social network of Figure~\ref{fig:illustrations}(d) is redrawn as a partial $k$-tree structure with $k=5$ in Figure~\ref{fig:illustrations}(e). Similarly, the 494 bus power system \cite{DavisACM:2011}, recently considered as a test network for loopy belief propagation \cite{Kirkley:2021}, is found to be a partial $10$-tree, using the method from ref.\ \cite{KlemmJPC:2020}. For these two and further empirical networks, we calculate the exact specific heat of the Ising model at varying temperature, see Figure~\ref{fig:heat}. For each choice of network and parameter $\beta$, the calculation takes less than one second on a single processor core of a standard portable computer.

%%% DISCUSSION
The present exact method is not restricted to the Ising model with all equal bonds but is equally efficient with any real-valued coupling strengths. These enter only in the initial factors of the computation. In spin glass models \cite{BinderRMP:1986} with frustrated Hamiltonians, ground state energies and freezing temperatures may be analyzed. We see potential to generalize the method to processes beyond the Ising model, such as bond and site percolation on networks \cite{RadicchiPRE:2015}. 

% Fast and exact evaluation of macroscopic quantities allows for many repetitions, e.g. for parameter scans or exploring the macroscopic dependence on weights assigned to the bonds.

%%%%%%%%%%%%%%%%%%%%%% METHODS %%%%%%%%%%%%%%%%%%%%%%%%%%%%%%%%%%%%%%%%%%%%%%%%%%%%%%%%%%%%%%%%%%%%%%%%%
\section*{Methods and Materials}
%%%%%%%%%%%%%%%%%%%%%%%%%%%%%%%%%%%%%%%%%%%%%%%%%%%%%%%%%%%%%%%%%%%%%%%%%%%%%%%%%%%%%%%%%%%%%%%%%%%%%%%%

\paragraph{Exact algorithm for the Ising partition function.} The algorithm takes as inputs a network given by a node set $V$ and bond set $B$, a value of inverse temperature $\beta$, and an ordering $(v(1),v(2),\dots,v(n)\}$ of $V$. Operations take place on a collection ${\mathcal F}$ of functions, called factors. Each factor $F \in {\mathcal F}$ has as its arguments one or several spin variables. Function values are real numbers. A factor can be implemented as a table or multidimensional array. Initially for each bond $\{v,w\}$ of the network, one factor with arguments $\sigma_v,\sigma_w$ and function values $F(\sigma_v,\sigma_w) = \exp(\beta \sigma_v \sigma_w)$ is contained in ${\mathcal F}$, cf.\ Equation (\ref{eq:ZwithProd}). $Z(\beta)$ is computed in a loop with index $i$ running from $1$ to $n$. For each $i$, the following four operations are performed. (I) form the product $F_{v(i)}$ of all factors in ${\mathcal F}$ that depend on spin $\sigma_{v(i)}$; (II) remove these factors from ${\mathcal F}$; (III) obtain $\hat F_{v(i)}$ by summing over $\sigma_v \in \{-1,+1\}$ in $F_{v(i)}$. (IV) include $\hat F_{v(i)}$ in ${\mathcal F}$. Upon completion of the loop, ${\mathcal F}$ contains a single factor being a single number $Z$, the result of the computation.

\paragraph{Derivatives of $Z$.} The specific heat
\begin{align}
C(\beta) & = \frac{\partial}{\partial T}\frac{\partial (-\ln Z(\beta))}{\partial \beta}\\
         & = \beta^2 \left[ Z(\beta)^{-1} \frac{\partial^2 Z}{\partial \beta^2} 
               Z(\beta)^{-2} \left( \frac{\partial Z}{\partial \beta} \right)^2 
                     \right]
\end{align}
involves the first and second derivatives of $Z(\beta)$, using $\partial / \partial T = - \beta^2 \partial / \partial \beta$. The exact values of these derivatives are computed within the same procedure as $Z$ itself. For the first derivative, a function $F'(\sigma_v,\sigma_w) = \sigma_v \sigma_w \exp(\beta \sigma_v \sigma_w)$ is generated initially for each bond $\{v,w\}$. When forming the product $F_\text{prod}$ of factors $F_1$ and $F_2$ in step (I) of the algorithm's loop, the derivative is obtained as
\begin{equation}
F_\text{prod}'(\{\sigma\}) = F_1'(\{\sigma\}) F_2(\{\sigma\}) + F_1(\{\sigma\}) F_2'(\{\sigma\})
\end{equation}
for each configuration $\{\sigma\}$ of spins these functions depend on.
Likewise the second derivative of the same product is
\begin{align}
F_\text{prod}''(\{\sigma\}) & = F_1''(\{\sigma\}) F_2(\{\sigma\}) + 2 F_1(\{\sigma\}) F_2'(\{\sigma\})\\
                           & + F_1 (\{\sigma\}) F_2''(\{\sigma\})~. \nonumber
\end{align}
We initialize $F''(\sigma_v,\sigma_w) = (\sigma_v \sigma_w)^2 \exp(\beta \sigma_v \sigma_w)$ for each bond $\{v,w\} \in B$ .

\paragraph{Models of growing 2-trees.} In all models, the initial condition is a network consisting of a single edge between two nodes. In the deterministic growth model by Dorogovtsev et al.\ \cite{DorogovtsevPRE:2002b}, the network of generation $g+1$ is obtained by adding a node $x$ and forming a triangle $(v,w,x)$ simultaneously for each bond $\{v,w\}$ present at generation $g$. The stochastic version of the model \cite{DorogovtsevPRE:2001} performs this node addition and triangle formation for one randomly selected bond in each microstep of growth. The {\em growth and deactivation model} by Klemm and Egu\'{\i}luz \cite{KlemmPRE:2002a} assigns each node a binary state as being active or inactive. In the initial condition, the two nodes present are active. At each step of growth, a new node $x$ forms a bond with each of the active nodes; $x$ is set active itself; out of the then three active nodes, one node is chosen randomly and deactivated. The probability of choosing a node $y$ for deactivation is proportional to $d_y^{-1}$, the inverse of the degree. If, instead, the oldest active node is deactivated in each step, a one-dimensional lattice with coordination number $4$ and open boundaries is obtained.

\paragraph{Elimination orders.} As empirical networks, we consider the 
karate club \cite{Zachary:1977}; social 3, protein 2, social 1 \cite{Milo:2004}; dolphins \cite{Lusseau:2003}, political books \cite{Krebs:2004}; s 838 \cite{Milo:2002}; network science \cite{Newman:2006}. For these networks (among others), suitable node orderings have been found by simulated annealing \cite{KlemmJPC:2020}, cf.\ the supplement of that article. For the 494 bus power system \cite{DavisACM:2011}, recently considered as a test network for loopy belief propagation \cite{Kirkley:2021}, a node ordering of maximum width 10 is found with the method of ref.\ \cite{KlemmJPC:2020} for all cooling schedules considered, even without any temperature variation. 

\section*{Acknowledgments}
Funding from MINECO through the Ram{\'o}n y Cajal program and through project SPASIMM, FIS2016-80067-P (AEI/FEDER, EU) is acknowledged.

\bibliographystyle{unsrt}
\bibliography{recising}

\begin{thebibliography}{10}

\bibitem{Ising:1925}
Ernst Ising.
\newblock Beitrag zur {Theorie} des {Ferromagnetismus}.
\newblock {\em Zeitschrift f\"{u}r Physik}, 31(1):253--258, 1925.

\bibitem{MaNJP:2019}
Yi-Ping Ma, Ivan Sudakov, Courtenay Strong, and Kenneth~M Golden.
\newblock Ising model for melt ponds on arctic sea ice.
\newblock {\em New Journal of Physics}, 21(6):063029, 2019.

\bibitem{SchneidmanNature:2006}
Elad Schneidman, Michael~J. Berry~II, Ronen Segev, and William Bialek.
\newblock Weak pairwise correlations imply strongly correlated network states
  in a neural population.
\newblock {\em Nature}, 440:1007--1012, 2006.

\bibitem{StaufferAJP:2008}
D.~Stauffer.
\newblock Social applications of two-dimensional {Ising} models.
\newblock {\em American Journal of Physics}, 76(4):470--473, 2008.

\bibitem{Bornholdt:2001}
Stefan Bornholdt.
\newblock Expectation bubbles in a spin model of markets: Intermittency from
  frustration across scales.
\newblock {\em International Journal of Modern Physics C}, 12(05):667--674,
  2001.

\bibitem{BaxterBook:1989}
Rodney~J. Baxter.
\newblock {\em Exactly Solved Models in Statistical Mechanics}.
\newblock Academic Press (London), 1989.

\bibitem{DorogovtsevPRE:2002}
S.~N. Dorogovtsev, A.~V. Goltsev, and J.~F.~F. Mendes.
\newblock Ising model on networks with an arbitrary distribution of
  connections.
\newblock {\em Phys. Rev. E}, 66:016104, 2002.

\bibitem{KlemmJPC:2020}
Konstantin Klemm.
\newblock Tree decompositions of real-world networks from simulated annealing.
\newblock {\em Journal of Physics: Complexity}, 1(3):035003, 2020.

\bibitem{MezardBook:2009}
Marc Mezard and Andrea Montanari.
\newblock {\em Information, Physics, and Computation}.
\newblock Oxford University Press, Inc., USA, 2009.

\bibitem{DorogovtsevRMP:2008}
S.~N. Dorogovtsev, A.~V. Goltsev, and J.~F.~F. Mendes.
\newblock Critical phenomena in complex networks.
\newblock {\em Rev. Mod. Phys.}, 80:1275--1335, 2008.

\bibitem{DemboAAP:2010}
Amir Dembo and Andrea Montanari.
\newblock {Ising models on locally tree-like graphs}.
\newblock {\em The Annals of Applied Probability}, 20(2):565 -- 592, 2010.

\bibitem{Watts:1998}
Duncan~J Watts and Steven~H Strogatz.
\newblock Collective dynamics of small-world networks.
\newblock {\em Nature}, 393(6684):440--442, 1998.

\bibitem{NewmanPNAS:2006}
M.~E.~J. Newman.
\newblock Modularity and community structure in networks.
\newblock {\em Proceedings of the National Academy of Sciences},
  103(23):8577--8582, 2006.

\bibitem{YedidiaNIPS:2000}
Jonathan~S. Yedidia, William~T. Freeman, and Yair Weiss.
\newblock Generalized belief propagation.
\newblock In {\em Advances in Neural Information Processing Systems 13},
  NIPS'00, pages 668--674, Cambridge, MA, USA, 2000. MIT Press.

\bibitem{RadicchiPRE:2016}
Filippo Radicchi and Claudio Castellano.
\newblock Beyond the locally treelike approximation for percolation on real
  networks.
\newblock {\em Phys. Rev. E}, 93:030302, 2016.

\bibitem{Cantwell:2019}
George~T. Cantwell and M.~E.~J. Newman.
\newblock Message passing on networks with loops.
\newblock {\em Proceedings of the National Academy of Sciences},
  116(47):23398--23403, 2019.

\bibitem{Kirkley:2021}
Alec Kirkley, George~T. Cantwell, and M.~E.~J. Newman.
\newblock Belief propagation for networks with loops.
\newblock {\em Science Advances}, 7(17), 2021.

\bibitem{Bodlaender:2010}
Hans~L. Bodlaender and Arie~M.C.A. Koster.
\newblock Treewidth computations i. upper bounds.
\newblock {\em Information and Computation}, 208(3):259--275, 2010.

\bibitem{KlemmPRE:2002a}
Konstantin Klemm and V\'{\i}ctor~M. Egu\'{\i}luz.
\newblock Highly clustered scale-free networks.
\newblock {\em Phys. Rev. E}, 65:036123, Feb 2002.

\bibitem{DorogovtsevPRE:2001}
S.~N. Dorogovtsev, J.~F.~F. Mendes, and A.~N. Samukhin.
\newblock Size-dependent degree distribution of a scale-free growing network.
\newblock {\em Phys. Rev. E}, 63:062101, 2001.

\bibitem{DorogovtsevPRE:2002b}
S.~N. Dorogovtsev, A.~V. Goltsev, and J.~F.~F. Mendes.
\newblock Pseudofractal scale-free web.
\newblock {\em Phys. Rev. E}, 65:066122, 2002.

\bibitem{DavisACM:2011}
Timothy~A. Davis and Yifan Hu.
\newblock The university of florida sparse matrix collection.
\newblock {\em ACM Trans. Math. Softw.}, 38(1), 2011.

\bibitem{BinderRMP:1986}
K.~Binder and A.~P. Young.
\newblock Spin glasses: Experimental facts, theoretical concepts, and open
  questions.
\newblock {\em Rev. Mod. Phys.}, 58:801--976, 1986.

\bibitem{RadicchiPRE:2015}
Filippo Radicchi.
\newblock Predicting percolation thresholds in networks.
\newblock {\em Phys. Rev. E}, 91:010801, 2015.

\bibitem{Zachary:1977}
Wayne~W. Zachary.
\newblock An information flow model for conflict and fission in small groups.
\newblock {\em Journal of Anthropological Research}, 33(4):452--473, 1977.

\bibitem{Milo:2004}
Ron Milo, Shalev Itzkovitz, Nadav Kashtan, Reuven Levitt, Shai Shen-Orr, Inbal
  Ayzenshtat, Michal Sheffer, and Uri Alon.
\newblock Superfamilies of evolved and designed networks.
\newblock {\em Science}, 303(5663):1538--1542, 2004.

\bibitem{Lusseau:2003}
D~Lusseau, K~Schneider, {O J} Boisseau, P~Haase, E~Slooten, and {S M} Dawson.
\newblock The bottlenose dolphin community of doubtful sound features a large
  proportion of long-lasting associations - can geographic isolation explain
  this unique trait?
\newblock {\em Behavioral Ecology and Sociobiology}, 54:396--405, 2003.

\bibitem{Krebs:2004}
Valdis Krebs, 2004.
\newblock Unpublished, data posted online at \url{http://www.orgnet.com/}.

\bibitem{Milo:2002}
R.~Milo, S.~Shen-Orr, S.~Itzkovitz, N.~Kashtan, D.~Chklovskii, and U.~Alon.
\newblock Network motifs: Simple building blocks of complex networks.
\newblock {\em Science}, 298(5594):824--827, 2002.

\bibitem{Newman:2006}
M.~E.~J. Newman.
\newblock Finding community structure in networks using the eigenvectors of
  matrices.
\newblock {\em Phys. Rev. E}, 74:036104, 2006.

\end{thebibliography}

\end{document}